\documentclass{article}

\usepackage[left=3cm, right=3cm, top=2cm, bottom = 2cm]{geometry}

\usepackage{graphicx}

\usepackage[xindy]{glossaries}

\newacronym{ltl}{LTL}{Linear-time Temporal Logic}

\newacronym{ctl}{CTL}{Computation Tree Logic}

\newacronym{ptl}{PTL}{Probabilistic Temporal Logic}

\newacronym{pctl}{PCTL}{Probabilistic Computation Tree Logic}

\newacronym{pctl*}{PCTL*}{Probabilistic Computation Tree Logic*}

\newacronym{tctl}{TCTL}{Timed Computation Tree Logic}

\newacronym{stl}{STL}{Signal Temporal Logic}

\newacronym{ltl-x}{LTL-X}{a version of Linear Time Logic without the `next' operator}

\newacronym{mtl}{MTL}{Metric Temporal Logic}

\newacronym{iactlk-x}{IACTLK-X}{a fragment of the temporal-epistemic logic CTLK, without the `next' operator}

\newacronym{mucalc}{$\mu$-calculus}{a strict superset of other temporal logics}

\newacronym{fotl}{FOTL}{First-Order Temporal Logic}

\newacronym{bdi}{BDI}{Belief-Desire-Intention}

\newacronym{prs}{PRS}{Procedural Reasoning System}

\newacronym{are}{ARE}{Autonomous Reasoning Engine}

\newacronym{dmars}{dMARS}{distributed Multi-Agent Reasoning System}

\newacronym{fsm}{FSM}{Finite-State Machine}

\newacronym{pfsm}{PFSM}{Probabilistic Finite-State Machine}

\newacronym{apfsm}{APFSM}{\textit{Autonomous} Probabilistic Finite-State Machine}

\newacronym{pha}{PHA}{Probabilistic Hybrid Automata}

\newacronym{fsa}{FSA}{Finite-State Automata}

\newacronym{ta}{TA}{Timed Automata}

\newacronym{gspn}{GSPN}{Generalised Stochastic Petri Net}

\newacronym{mopn}{MOPN}{Marked Ordinary Petri Net}

\newacronym{ba}{BA}{B\"{u}chi Automata}

\newacronym{dtmc}{DTMC}{Discrete-Time Markov Chain}

\newacronym{pars}{PARS}{Process Algebra for Robot Schemas}

\newacronym{fsp}{FSP}{Finite State Processes}

\newacronym{piadl}{$\pi$ADL}{$\pi$ calculus combined with ADL}

\newacronym{csp}{CSP}{Communicating Sequential Processes}

\newacronym{ccs}{CCS}{Calculus of Communicating Systems}

\newacronym{wsccs}{WSCCS}{Weighted Synchronous CCS}

\newacronym{csp|b}{CSP$\|$B}{an integrated formalism combining CSP and B}

\newacronym{fdr}{FDR}{the Failures-Divergences Refinement checker}

\newacronym{jpf}{JPF}{Java PathFinder}

\newacronym{ajpf}{AJPF}{Agent Java PathFinder}

\newacronym{mcmas}{MCMAS}{Model Checker for Multi-Agent Systems}

\newacronym{ltlmop}{LTLMoP}{Linear Temporal Logic MissiOn Planning}

\newacronym{adl}{ADL}{Architecture Description Language}

\newacronym{aadl}{AADL}{Architecture Analysis and Design Language}

\newacronym{lts}{LTS}{Labelled-Transition System}

\newacronym{uml}{UML}{Unified Modelling Language}

\newacronym{sysml}{SySML}{Systems Modelling Language}

\newacronym{robotml}{RobotML}{Robot Modelling Language}

\newacronym{dsl}{DSL}{Domain Specific Language}

\newacronym{sct}{SCT}{Supervisory Control Theory}

\newacronym{ide}{IDE}{Integrated Development Environment}

\newacronym{mde}{MDE}{Model-Driven Engineering}

\newacronym{ai}{AI}{Artificial Intelligence}

\newacronym{fdir}{FDIR}{Fault Detection, Isolation and Recovery}

\newacronym{ifm}{iFM}{Integrated Formal Methods}

\newacronym{ros}{ROS}{Robot Operating System}

\newacronym{amcl}{AMCL}{Adaptive Monte Carlo Localisation}

\newacronym{dl}{\text{\upshape\textsf{d{\kern-0.05em}L}}}{differential dynamic logic}

\newacronym{vm}{VM}{Virtual Machine}

\newacronym{bip}{BIP}{Behaviour Interaction Priority}

\newacronym{fol}{FOL}{First-Order Logic}

\newacronym{owl}{OWL}{Web Ontology Language}

\newacronym{ria}{RIA}{Restore Invariant Approach}

\newacronym{ocl}{OCL}{Object Constraint Language}

\newacronym{sil}{SIL}{Safety Integrity Level}

\newacronym{rcl}{RCL}{ROS Contract Language}

\newacronym{rv}{RV}{Runtime Verification}

\newacronym{rml}{RML}{Runtime Monitoring Language}
\usepackage[hidelinks]{hyperref}
\usepackage{doi}
\usepackage{multirow}
\usepackage{paralist}
\usepackage{textcomp}
\usepackage[utf8]{inputenc}
\usepackage[T1]{fontenc}
\usepackage{lmodern}
\usepackage[UKenglish]{isodate}

\usepackage{amsmath}
\usepackage{xcolor}

\usepackage{cite}

\usepackage{scalerel}
\usepackage{tikz}
\usetikzlibrary{svg.path}

\definecolor{orcidlogocol}{HTML}{A6CE39}
\tikzset{
  orcidlogo/.pic={
    \fill[orcidlogocol] svg{M256,128c0,70.7-57.3,128-128,128C57.3,256,0,198.7,0,128C0,57.3,57.3,0,128,0C198.7,0,256,57.3,256,128z};
    \fill[white] svg{M86.3,186.2H70.9V79.1h15.4v48.4V186.2z}
                 svg{M108.9,79.1h41.6c39.6,0,57,28.3,57,53.6c0,27.5-21.5,53.6-56.8,53.6h-41.8V79.1z M124.3,172.4h24.5c34.9,0,42.9-26.5,42.9-39.7c0-21.5-13.7-39.7-43.7-39.7h-23.7V172.4z}
                 svg{M88.7,56.8c0,5.5-4.5,10.1-10.1,10.1c-5.6,0-10.1-4.6-10.1-10.1c0-5.6,4.5-10.1,10.1-10.1C84.2,46.7,88.7,51.3,88.7,56.8z};
  }
}

\newcommand\orcidicon[1]{\href{https://orcid.org/#1}{\mbox{\scalerel*{
\begin{tikzpicture}[yscale=-1,transform shape]
\pic{orcidlogo};
\end{tikzpicture}
}{|}}}}

\newcommand{\tnbs}{\hspace{0.4\smallskipamount}}
\renewenvironment{quote}
  {
  \small\list{}{\setlength\topsep{2pt}\rightmargin=0.2cm \leftmargin=0.2cm}%
   \item\relax}
  {\endlist}
 
\let\oldcenter\center
\let\oldendcenter\endcenter
\renewenvironment{center}{\setlength\topsep{1pt}\oldcenter}{\oldendcenter}

\usepackage{tabularx}
\usepackage{wrapfig}
\usepackage{fretish} 

\usepackage{tikz}
\usetikzlibrary{shapes.geometric, arrows}
\tikzstyle{fragment} = [rectangle, minimum width=10em, minimum height=2em, text centered, draw=black, fill={rgb:black,1;white,3}]
\tikzstyle{req} = [rectangle, minimum width=2em, minimum height=2em, text centered, draw=black]
\tikzstyle{arrow} = [thick,->,>=stealth]

\usepackage[]{todonotes}

\begin{document}

\title{Why just FRET when you can Refactor?\\ Retuning FRETISH Requirements
\thanks{This research was funded by the European Union’s Horizon 2020 research and innovation programme under the VALU3S project (grant No 876852), and by Enterprise Ireland (grant No IR20200054). The funders had no role in study design, data collection and analysis, decision to publish, or preparation of the manuscript.}}


\author{Matt Luckcuck \and Marie Farrell \and Ois\'{i}n Sheridan}




\date{Department of Computer Science, Maynooth University, Ireland\\
\texttt{valu3s@mu.ie}}

\maketitle             

\begin{abstract}
 
Formal verification of a software system relies on formalising the requirements to which it should adhere, which can be challenging. While formalising requirements from natural-language, we have dependencies that lead to duplication of information across many requirements, meaning that a change to one requirement causes updates in several places. We propose to adapt code refactorings for NASA’s \gls{fret}, our tool-of-choice. Refactoring is the process of reorganising software to improve its internal structure without altering its external behaviour; it can also be applied to requirements, to make them more manageable by reducing repetition. \gls{fret} automatically translates requirements (written in its input language \fretish{}) into Temporal Logic, which enables us to formally verify that refactoring has preserved the requirements' underlying meaning. In this paper, we present four refactorings for \fretish{} requirements and explain their utility. We describe the application of one of these refactorings to the requirements of a civilian aircraft engine software controller, to decouple the dependencies from the duplication, and analyse how this changes the number of requirements and the number of repetitions. We evaluate our approach using Spot, a tool for checking equivalence of Temporal Logic specifications.

\end{abstract}

\glsresetall

\section{Introduction}
\label{sec:intro}


Developing a valid set of system requirements necessitates discussion with people who have expertise in the system under development, who may not be experts in formal methods. Requirements elicitation discussions can be helped by writing
the requirements in an intermediate, semi-formal language before fully formalising them.
We use NASA's \gls{fret} as a gateway for developing formal requirements alongside an industrial partner. \gls{fret} integrates a semi-formal requirements language and temporal logic \cite{giannakopoulou_formal_2020}.

\textit{Refactoring} is applied to software to improve its structure and is defined as:
\begin{quote}
\textit{``...the process of changing a software system in such a way that it does not alter the external behaviour of the code yet improves its internal structure''}~\cite{fowler_refactoring_1999}.
\end{quote}
The cleaner code produced by refactoring is easier to maintain, examine, understand, and update.
Like software, requirements often go through several iterations before they are complete. Even then, they may need updating, if an error is found or a new feature is added. This means that their structure is almost 
as important as that of the software that they specify.

While eliciting the requirements for a civilian aircraft engine software controller, as part of the VALU3S project\footnote{The VALU3S project: \url{https://valu3s.eu/}}, we found dependencies that lead to many repeated definitions, meaning that one change required updates in multiple places -- a classic \textit{bad smell} in software engineering~\cite{fowler_refactoring_1999} -- which makes the requirements much more difficult to maintain. 

In this paper, we adapt code refactorings to requirements written in \gls{fret}'s input language, \fretish{}. We then present our refactoring of the requirements for our example application. Our industrial partner supplied 14 abstract requirements, which we mapped one-to-one into \fretish{}. Elicitation discussions with our industrial partner produced 42 \fretish{} requirements in total~\cite{farrell2021fretting}. 

Our refactoring approach was driven by the need to reorganise the requirements set. Refactoring the requirements slightly reduced the total number of requirements, and markedly reduced the number of repeated definitions. We check  that  the  refactoring  steps  have  not  changed  the behaviour of the requirements by using an external library to compare their temporal logic translations.
Our previous work explores the motivation for refactoring \fretish{} requirements~\cite{farrell2022refactoring}, and in this paper we introduce the refactoring steps and present a worked example.

This paper is laid out as follows. \S\ref{sec:bg} presents the background of \gls{fret} (\S\ref{sec:fretIntro}) and requirements refactoring (\S\ref{sec:refactoringIntro}). \S\ref{sec:refactoring} describes our adaptation of code refactorings to \fretish{} requirements, and \S\ref{sec:example} describes how we refactored our requirements set. In \S\ref{sec:analysis}, we analyse how refactoring has changed our requirements set. \S\ref{sec:discussion} discusses how to introduce refactoring into \gls{fret} and what impact that might have. Finally, \S\ref{sec:conclusion} concludes the paper.

\section{Background}
\label{sec:bg}

This section provides an overview of \gls{fret} and refactoring for requirements.

\subsection{FRET}
\label{sec:fretIntro}

\gls{fret}\footnote{FRET: \url{https://github.com/NASA-SW-VnV/fret}} is an open-source tool that enables developers to write and formalise system requirements in a structured natural-language called \fretish{}~\cite{giannakopoulou_formal_2020}. Requirements in \gls{fret} take the form:\\ 
\centerline{\fretishComponents{}}
\begin{wrapfigure}[14]{l}{0.6\textwidth}
\centering
\includegraphics[width= 0.6\textwidth]{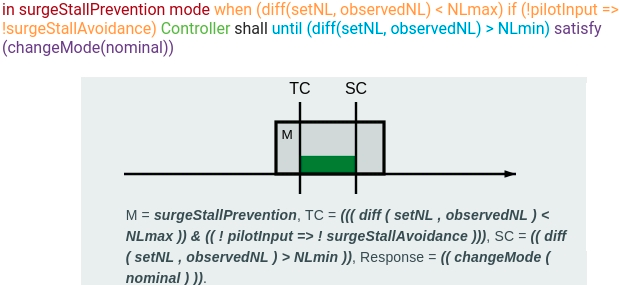}
\caption{Diagramatic Semantics for one of our requirements. M indicates the \Scope{} of the requirement, TC denotes the Triggering Condition and SC the stopping condition. \label{fig:diagramaticSemantics}}
\end{wrapfigure}
\noindent Here, the \Scope{} and \Timing{} fields are optional. Intuitively, this structure defines requirements that, in a particular \Scope{} and under a given \Condition{} then a particular \Component{} shall obey the \Response{} under specific \Timing{} constraints.

\gls{fret} 
formalises each requirement in both past- and future-time \gls{mtl}.  
For each requirement, \gls{fret} also produces a structured text description of the requirement's behaviour, and a diagramatic semantics (Fig.\ref{fig:diagramaticSemantics}) that shows the time interval where the requirement should hold and the requirement's triggering and stopping conditions (if they exist). These are helpful for sanity-checking what has been written in \fretish{}.

\gls{fret} supports a hierarchical link between requirements. In this many-to-many relationship, \textit{parent} requirements may have many children, and a \textit{child} requirement may have many parents. Linking parent and child requirements based on their IDs supports traceability, but the parent-child relationship does not include inheritance or any other functional link between the requirements.

\gls{fret} can automatically translate requirements into contracts for a Simulink diagram, written in CoCoSpec, which are checked during Simulink simulations by the CoCoSim tool~\cite{bourbouh2020cocosim}. \gls{fret} can also generate runtime monitors for the stream-based Copilot framework~\cite{dutle2020requirements}.

\subsection{Refactoring Requirements}
\label{sec:refactoringIntro}

Refactoring is a software engineering process where program code is reorganised to improve its internal structure, without altering its external behaviour~\cite{fowler_refactoring_1999}. Refactoring has been adapted for software requirements, for example Ramos et al.~\cite{ramos_improving_2007} present refactoring approaches for requirements, directly based on the code refactorings described by Fowler et al.~\cite{fowler_refactoring_1999}. Both of these works have been particularly influential on this paper.

Requirements are an abstraction of a system's model which are, in turn, abstractions of its implementation.
Refactoring can also be applied to (formal) system models~\cite{correa2007refactoring,liu2008refactoring,gheyi2004refactoring}. Refactoring a system's requirements has the potential to avoid modelling or implementing problems that already exist in the requirements. 
Some code refactorings may weaken traceability links back to the requirements~\cite{mahmoud_supporting_2014}. 
Weakened traceability was shown when extracting code into a method, but moving a method between classes had no effect on traceability; renaming software units had a positive effect on traceability. 

Being able to refactor requirements enables us to deal with the, almost inevitable, likelihood that requirements will need to change~\cite{deshpande2019data}. This can be because of new features being added to the system, or requirements elicitation discussions identifying new requirements. 
For our requirements set, discussions with our industrial partner identified concrete details about the requirements, but
compounded the problem of duplication: with many definitions, one change required updates in many requirements. We discuss this in detail in \S\ref{sec:example}.

Deshpande et al. found that identifying dependencies between requirements is important, and that ignoring them can negatively impact a project's success~\cite{deshpande2019data}. 
They used machine learning to identify dependencies between natural-language requirements in the same set, including \textsc{requires} and \textsc{similar} relationships. 
Their survey of practitioners revealed that, although important, over 90\% of participants did not use an automated tool to extract and maintain dependencies. In \S\ref{sec:step1}, we manually identify  dependencies within our own requirements set. As mentioned, \gls{fret} captures hierarchical links between parent- and child-requirements, but does not yet enable requirement dependencies to be captured, apart from textually as part of a requirement's rationale.


\gls{fret} automatically generates a formal representation in temporal logic, which enables us to formally verify that the refactoring has not changed the requirements' behaviour. This is analogous to the crucial `compile and test' step for code refactoring~\cite{fowler_refactoring_1999}. 
This step is not possible in~\cite{ramos_improving_2007}, because they focus on textual requirements so there is no easy way to test that the meaning of the requirements have not changed.

\section{Refactoring FRETISH: Concepts}
\label{sec:refactoring}

\gls{fret} translates \fretish{} requirements into (past- and future-time metric) temporal logic, but the \fretish{} requirements themselves are simply structured text. This means that refactoring \fretish{} is very similar to refactoring natural-language requirements. This section describes the adaptation of four of the requirements refactorings from~\cite{ramos_improving_2007}, to \fretish{} requirements. The fifth refactoring in~\cite{ramos_improving_2007} (\textsc{Extract Alternative Flows}) is not applicable to \fretish{}. 

We propose that tool-supported refactoring should be included in \gls{fret}, which should include a formal check that the temporal logic version of the requirements before the refactoring is the same as after the refactoring (i.e. that the behaviour has not been changed). This can be achieved by checking that the two temporal logic formulae are equivalent; or that the refactored requirement implies the original, if the refactoring adds behaviour to the requirement but still includes the behaviours of the original.

Even with tool support, refactoring is not a process that should be applied through blind automation \cite{ramos_improving_2007}. Users who are formalising requirements must decide which refactorings are appropriate for their set of requirements. 




\subsection{Extract Requirement}
\label{sec:extract}

\textsc{Extract Requirement} moves definitions from one requirement into a newly-created requirement. In the original requirement, the extracted parts are replaced with a reference to the new requirement. This is an adaptation of \textsc{Extract Requirement} in~\cite{ramos_improving_2007} (itself based on \textsc{Extract Method} from~\cite{fowler_refactoring_1999}) to \fretish{} requirements. When refactoring software, \textsc{Extract Method} extracts a piece of repeated code into its own method. \textsc{Extract Requirement} achieves similar modularity for \fretish{} requirements.  
\textsc{Extract Requirement} also serves the intent of the \textsc{Extract Alternative Flows} refactoring~\cite{ramos_improving_2007}, which is aimed at modularising Use Case descriptions and has no clear application in \fretish{}.

Using \textsc{Extract Requirement} to introduce a reference to the newly-created requirement can improve readability, because it can be named to better inform the reader of its intent. This refactoring is especially useful when the definition being extracted is repeated across several requirements, because encapsulating it inside one requirement means that a change to the behaviour only requires updates in one place. To perform the \textsc{Extract Requirement} refactoring in \gls{fret}, we propose the following steps:
\begin{compactitem}
\item[\textbf{Step 1}] Create a new requirement with a descriptive name for the definitions being extracted;
\item[\textbf{Step 2}] Copy the definitions being extracted to the newly-created requirement;
\item[\textbf{Step 3}] Replace the extracted definitions in the original requirement with a reference to the newly-created requirement;
\item[\textbf{Step 4}] Verify that the behaviour of the original requirement combined with the newly-created requirement is the same as (or at least implies) the behaviour of the original requirement before refactoring.
\end{compactitem}
\noindent Each extracted requirement follows the same pattern: the extracted definitions become its \Condition{}, and its \Response{} sets a boolean to $True$ if the \Condition{} holds. This represents the extracted definition being satisfied. 
The original requirement's reference to the extracted requirement is interpreted by \gls{fret} as a boolean variable. Since we know that the extracted requirement is $True$ if the extracted definitions are $True$, we can intuitively say that the reference in the original requirement should only be $True$ if the extracted conditions are $True$. 

This intuition is analogous to having extracted a complicated boolean condition into a new method that returns a boolean, then replacing the original condition with a call to the method. To capture this link in \gls{fret}, the `Rationale' field of both the original and extracted requirements could be used to describe how refactoring has changed the requirements. \gls{fret} does not currently support `calling' methods and \S\ref{sec:discussion} outlines ways of adding this support.\smallskip
 
\noindent\textit{Example:}
Consider a requirement, \textsf{R1}, for a sensor subsystem component. \textsf{R1} states that ``\textit{if} $sensorA$ \textit{returns a value outside of its expected range, plus a margin, then it is invalid}'', (\response{!sensorA\_valid}). In \fretish{}, \textsf{R1} becomes:\\
\centerline{ \condition{if (sensorA $>$ sensorA\_Max $+$ R)\tnbs{}$|$(sensorA $<$ sensorA\_Min $-$ R)}} 
\centerline{ \component{SensorSubSystem} \texttt{shall} \response{!sensorA\_valid}} 

\noindent Applying \textsc{Extract Requirement} to the condition \condition{sensorA\tnbs{}$>$\tnbs{}sensorA\_Max\tnbs{}+\tnbs{}R)\tnbs{}|\tnbs{}(sensorA\tnbs{}$<$\tnbs{}sensorA\_Min\tnbs{}-\tnbs{}R)} moves these definitions to their own requirement which we call \textsf{SENSOR\_A\_OUT\_OF\_BOUNDS}:\smallskip

\centerline{\condition{if (sensorA $>$ sensorA\_Max + R) | (sensorA $<$ sensorA\_Min - R) } }
\centerline{\component{SensorA} \texttt{shall} \response{ SensorA\_OutOfBounds} }

\smallskip

\noindent \textsf{R1} is updated to read:\\
\centerline{\condition{if \textsf{SENSOR\_A\_OUT\_OF\_BOUNDS}} \component{SensorSubSystem} \texttt{shall} \response{!sensorA\_valid} }

This refactoring means that any changes to the extracted conditions only require updates to \textsf{SENSOR\_A\_OUT\_OF\_BOUNDS}. It also simplifies \textsf{R1} and makes it easier to read, with the extracted requirement describing what triggers the response (\response{!sensorA\_valid}). Other repetitions of the condition can be similarly replaced by the extracted requirement. In each case a check is performed to ensure that the behaviour is preserved.


\subsection{Rename Requirement}

\textsc{Rename Requirement} changes a requirement's name and then updates any references to it to match its new name. \gls{fret} enables the user to rename a requirement, but any references to it in the `Parent Requirement ID' field of child requirements are not updated. This means that child requirements are left with a broken reference, reducing traceability. Updating the renaming function to act like \textsc{Rename Requirement} would solve this problem. It would also be useful when updating a requirement name that is referenced in many requirements.

In addition to the steps proposed by Ramos et al.~\cite{ramos_improving_2007}, this refactoring should also check that renaming has not caused any inconsistencies when applied over multiple requirements. Similar rename refactoring has been provided for other formal methods, e.g. Event-B\footnote{\url{https://wiki.event-b.org/index.php/Refactoring_Framework}}.

\subsection{Move Definition}

\textsc{Move Definition} takes part of a requirement and moves it to another, adapting \textsc{Move Activity}~\cite{ramos_improving_2007}.
It focusses a requirement on a single responsibility. Because this refactoring will change the behaviour of both the source and destination requirements, it should check that the parts that were moved are no longer in the source requirement, but are in the destination requirement. 

\gls{fret}'s parent-child link creates a requirements hierarchy, so \textsc{Move Definition} could be specialised into a \textsc{Pull-Up Definition} refactoring that explicitly moves common definitions from a child requirement up to its parent to represent the notion that the definitions is common to all of that requirement's children. This is similar to the \textsc{Pull-Up Method} code refactoring in~\cite{fowler_refactoring_1999}.

The parent-child link between \fretish{} requirements is based on their IDs, but it does not imply any semantic link between the requirements. However, moving a definition up the hierarchy makes it more obvious that it is common to all of the requirement's children. This is similar to a method being placed in the superclass of an Object-Oriented program's class hierarchy. Using \textsc{Pull-Up Definition} makes it more likely that this notion will be implemented.

\textsc{Pull-Up Definition} should check that the destination requirement \textit{implies} the source requirement, because the destination (parent) now covers the definitions in the source (child). When applying this refactoring, care must be taken not to reduce traceability of the origin or the pulled-up definition(s).
\smallskip

\noindent\textit{Example:} Consider the previous example of \textsf{R1} (\S\ref{sec:extract}). Elicitation discussions then reveal a new condition, ``\textit{switchA is triggered}'' (\condition{switchA = true}), which is added to \textsf{R1}'s children. Using \textsc{Pull-Up Definition} to move this condition from the children to \textsf{R1} simplifies the requirements set, and makes it clear that the condition is common to all of \textsf{R1}'s children. However, traceability is potentially reduced unless the condition's origin and destination are documented.

\subsection{Inline Requirement}
\label{sec:inline}
\textsc{Inline Requirement} is the opposite of the \textsc{Extract Requirement} refactoring (\S\ref{sec:extract}); it takes one requirement and merges it into another. It is useful where a requirement is only referenced in a few places, and is simple enough that its definitions are as clear as its name. 
Care must be taken when merging a  \fretish{} requirement's fields. 
This is relatively simple for the \Condition{} and \Response{} fields; but if the \Scope{}, \Component{}, or \Timing{} fields do not match, then \textsc{Inline Requirement} might not be applicable. (We discuss this further in \S\ref{sec:discussion}.)
\textsc{Inline Requirement} should check for equivalence between the original two requirements and the refactored requirement, similarly to \textsc{Extract Requirement} (\S\ref{sec:extract}).


\section{Refactoring FRETISH: Application}
\label{sec:example}

In this section we describe the application of the \textsc{Extract Requirement} refactoring (\S\ref{sec:extract}) to a set of \fretish{} requirements. Here, we focus on the whole requirement set; first identifying the parts of the requirements to refactor, then systematically extracting repeated definitions, before cleaning up the requirements set, and comparing the requirements to check that their behaviour has not changed. Because \gls{fret} does not automatically perform the comparison that we perform, and because we are refactoring the whole requirement set, we defer the comparison until \textit{after} all of the refactorings are complete.

\begin{sloppypar}
Our \fretish{} requirements are derived from the natural-language requirements for a a high-bypass civilian aircraft turbofan engine software controller, provided by our industrial partner on the VALU3S~\cite{barbosa2020valu3s} project, described in our prior work~\cite{luckcuck2021verifiable}.
The controller's high-level objectives are: to manage engine thrust, regulate compressor pressure and speeds, and limit engine parameters to safe values. It should continue to operate, keeping settling time, overshoot, and steady state errors within acceptable limits; while respecting the engine's operating limits in the presence of: sensor faults, perturbation of system parameters, and other low-probability hazards.
\noindent The controller must also detect engine surge/stall, and change mode to avoid these hazardous situations. 
\end{sloppypar}

Our \fretish{} requirements set is derived from the 14 English-language requirements and 20 abstract test cases provided by our industrial partner, see~\cite{farrell2021fretting}. We reuse the natural-language requirements' naming convention  in the \fretish{} set. 
As mentioned in \S\ref{sec:fretIntro}, \gls{fret} allows requirements to be hierarchically linked: a requirement can have many \textit{parent} requirements, and a parent requirement can have many \textit{child} requirements. 
Thus, the naming convention in our requirements: \\
\centerline{$<$\textsf{use case id}$>$\textsf{\_R\_}$<$\textsf{parent requirement id}$>$.$<$\textsf{child requirement id}$>$}\\
For example, this is Use Case 5 in the VALU3S project, so Requirement 1 is named \textsf{UC5\_R\_1}; we use a \textsf{sans serif} font style to denote requirement names. 

The natural-language requirements (examples in Table~\ref{tab:reqs123}) repeat concepts like \textit{``sensor faults''} or \textit{``control objectives''}, which we call \textsl{fragments}; we use a \textsl{slanted} font style to denote fragment names.
We mapped the natural-language requirements, one-to-one, in to the 14 \fretish{} parent requirements; in which the fragments are represented by boolean variables.
For example, the natural-language version of \textsf{UC5\_R\_1} shown in Table \ref{tab:reqs123} is encoded in \fretish{} as:\\
\centerline{\condition{if sensorFaults \& trackingPilotCommands} \component{Controller}}\\
\centerline{\texttt{shall} \response{controlObjectives}}
\noindent A \fretish{} \Condition{} can start with one of several different keywords, including \condition{if} and \condition{when}, which each have the same semantics. Here the \Condition{} captures the fragments as booleans, e.g.\textit{``Under sensor faults''} becomes \condition{sensorFaults}.

The child requirements add detail from the test cases and elicitation discussions with our industrial partner.
Each fragment's definition is duplicated	 in one or more child requirements. For example, \textsf{UC5\_R\_1} relies on its child requirements to define \condition{sensorFaults}, \condition{trackingPilotCommands}, and \response*{controlObjectives}.

\begin{table}[t]
\centering
\begin{tabular}{c|p{10.5cm}}
\textbf{ID} & \textbf{Description}\\ 
\hline \hline
\textsf{UC5\_R\_1} & Under sensor faults, while tracking pilot commands, control objectives shall be satisfied (e.g., settling time, 
overshoot, and steady state error will be within predefined, acceptable limits) \\ \hline
\textsf{UC5\_R\_3} & Under sensor faults, while tracking pilot commands, operating limit objectives shall be satisfied (e.g., respecting 
upper limit in shaft speed) \\ \hline
\end{tabular}
\caption{Natural-language requirements \textsf{UC5\_R\_1} and \textsf{UC5\_R\_3} for our case study. \label{tab:reqs123} }
\end{table}

This duplication of fragments leads to a lot of repetition.
For example, \textsl{Sensor Faults} appears in 4 parent requirements, with its definition duplicated in a further 8 children. 
This repetition is the \textit{Duplicated Activities} problem~\cite{ramos_improving_2007} (similar to the smell of \textit{Duplicated Code}~\cite{fowler_refactoring_1999}) meaning that a change to one fragment requires changes to all of the child requirements that use it.
The rest of this section describes how we identify the fragments in our requirements set, and reduce the repetition with the \textsc{Extract Requirement} refactoring (\S\ref{sec:extract}).

\subsection{Step 1: Identify Fragments}
\label{sec:step1}

This step identifies the repeated fragments that are `baked into' the requirements, so that we can apply \textsc{Extract Requirement} to them. 
We were able to perform this analysis manually, because our requirements set is relatively small. For larger requirements sets, automatic tools may be needed. For example using Machine Learning to identify dependencies within requirements sets~\cite{deshpande2019data}. 
We focus on two requirements, providing a useful illustration of our approach. 

Table~\ref{tab:reqs123} shows the natural-language version of \textsf{UC5\_R\_1} and \textsf{UC5\_R\_3}, provided by our industrial partner.
Both requirements are active \textit{``Under sensor faults''} and \textit{``while tracking pilot commands''}. Whereas \textsf{UC5\_R\_1} satisfies \textit{``control objectives''}, while \textsf{UC5\_R\_3} satisfies \textit{``operating limit objectives''}.

\begin{table}[t]
\centering
\begin{tabular}{c|p{7cm}}
\textbf{Fragment ID} & \textbf{Description} \\
\hline
\hline
F1 & \textsl{Sensor Faults} \\ \hline
F2 & \textsl{Tracking Pilot Commands} \\ \hline
F3 & \textsl{Control Objectives} \\ \hline
F4 & \textsl{Regulation Of Nominal Operation} \\ \hline
F5 & \textsl{Operating Limit Objectives} \\ \hline
F6 & \textsl{Mechanical Fatigue} \\ \hline
F7 & \textsl{Low Probability Hazardous Events} \\ \hline

\end{tabular}
\caption{The seven repeated fragments in the natural-language requirements. \label{tab:fragments}}
\end{table}

To analyse the whole requirements set, we first identify the fragments that are repeated in the natural-language requirements, which we list in Table \ref{tab:fragments}. We collect the  requirements and the fragments into two sets: \\
\centerline{$Reqs = \{ \textsf{UC5\_R\_1}, \dots, \textsf{UC5\_R\_14} \}$ and $Frags = \{F1, F2, \dots, F6, F7 \}$,}
which are linked by dependency: understanding a requirement is dependant on first understanding the fragments that it contains, a change to the fragment is duplicated across all of the requirements that depend on it. 

We capture the dependencies as a set of tuples: \\
\centerline{$ Dependencies = \{ (\textsf{UC5\_R\_1}, F1), (\textsf{UC5\_R\_1}, F2), (\textsf{UC5\_R\_1}, F3), \dots \}$}
\noindent each tuple maps a requirement to one fragment upon which it depends. 
There are 38 dependencies in the natural-language requirements, which are mapped into the \fretish{} parent requirements; a full dependency graph is shown in \cite{farrell2022refactoring}.

\begin{figure*}[t]
\centering
\scalebox{0.7}{
\begin{tikzpicture}[node distance=1.5em]

\node (SF) [fragment, xshift = -50em]  {\parbox[t][][t]{3.2cm}{\centering F1: \textsl{SensorFaults}}};

\node (TPC) [fragment, right of = SF, xshift = 12em]{\centering{F2: \textsl{TrackingPilotCommands}}};

\node (CO) [fragment, right of = TPC, xshift = 12em]{\parbox[t][][t]{3.2cm}{\centering F3: \textsl{ControlObjectives}}};

\node (OLO) [fragment, right of = CO, xshift = 12em]{\centering F5: \textsl{OperatingLimitObjectives}};

\node (R1) [req, circle, below of = SF, yshift = -3.5em, xshift = 10em]{\parbox[t][][t]{1.2cm}{\centering{\textsf{UC5\_R\_1}}}};

\node (R3) [req, circle, below of = CO, yshift = -3.5em, xshift = 5em]{\parbox[t][][t]{1.2cm}{\centering{\textsf{UC5\_R\_3}}}};

\draw [arrow ] (R1)  -- (TPC);
\draw [arrow] (R3)  -- (TPC);

\draw [arrow] (R1)  -- (SF);
\draw [arrow] (R3)  -- (SF);

\draw [arrow] (R1)  -- (CO);

\draw [arrow] (R3)  -- (OLO);

\end{tikzpicture}

}
\caption{The dependency graph between the requirements \textsf{UC5\_R\_1} and \textsf{UC5\_R\_3} (white circles) and the associated fragments (grey boxes). Arrows indicate a `depends on' relationship.}
\label{fig:r1r3Dependancy}
\end{figure*}
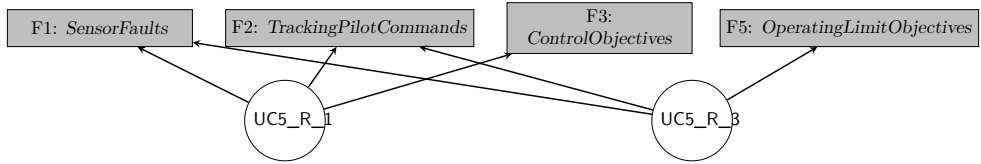

For brevity, we focus on the dependencies of \textsf{UC5\_R\_1} and \textsf{UC5\_R\_3}, 
shown in the dependency graph in Fig.~\ref{fig:r1r3Dependancy}. This figure exposes the mapping between the parent requirements and the fragments on which they depend. These dependencies are hidden within the requirements, which makes maintaining the duplicated definitions of the fragments troublesome.

Our elicitation discussions identified two more fragments within the child requirements, 
which define when a requirement is \textsl{Active} and \textsl{Not Active} by comparing the set and measured values of the system's variables. 
Each child requirement depends on both of these fragments, therefore we also extract them in Step 2 (\S\ref{sec:step2}).

In this original requirements set, dependencies always lead to duplication. 
In total, there are 94 dependencies, across the 42 requirements. This includes the 38 dependencies from by the natural-language requirements, and the 56 dependencies from the \textsl{Active} and \textsl{Not Active} fragments. 
We use this dependency information as the starting point for extracting the fragments. 

\subsection{Step 2: Apply Extract Requirement}
\label{sec:step2}

In this step we use the \textsc{Extract Requirement} (\S\ref{sec:extract}) refactoring to decouple the requirements from the duplicated fragments. 
We use \textsc{Extract Requirement} on the seven natural-language fragments (Table~\ref{tab:fragments}) and both the \textsl{Active} and \textsl{Not Active} fragments (\S\ref{sec:step1}) (which can be seen in the \condition{when()} and \timing{until()} clauses of each child requirement). 
In total, this step produces nine new requirements, each is the destination for an extracted definition.

We focus on \textsf{UC5\_R\_1} (Table~\ref{tab:reqs123}), which contains three fragments that are drawn from the natural-language version of the requirement: \condition{sensorFaults}, \condition{tra\-ckingPilotCommands}, and \response*{controlObjectives}. \textsf{UC5\_R\_1} has three child requirements. For example, \textsf{UC5\_R\_1.1} contains the detailed definitions of the three fragments from \textsf{UC5\_R\_1} and both of the \textsl{Active} and \textsl{Not Active} fragments:\\
\centerline{\condition{when\tnbs{}(diff(r(i),y(i))\tnbs{}\textgreater{}\tnbs{}E) if((sensorValue(S)\tnbs{}\textgreater{}\tnbs{}nominalValue+R) $|$}}
\centerline{\condition{(sensorValue(S)\tnbs{}\textless{}\tnbs{}nominalValue-R)\tnbs{}$|$\tnbs{}(sensorValue(S)=null) \&}}
\centerline{\condition{(pilotInput\tnbs{}=\textgreater{}\tnbs{}setThrust=V2)\tnbs{}\&\tnbs{}(observedThrust=V1))}}
\centerline{\component{Controller} \texttt{shall} \timing{until\tnbs{}(diff(r(i),y(i))\tnbs{}\textless{}\tnbs{}e)} \response{(settlingTime\tnbs{}\textgreater{}=\tnbs{}0) \&}}
\centerline{\response*{\tnbs{}(settlingTime\tnbs{}\textless{}=\tnbs{}settlingTimeMax)\tnbs{}\&\tnbs{}(observedThrust=V2)} }

\noindent The \textsl{Active} fragment is defined in the \condition{when()} clause of the \Condition{}. The first three comparisons in the \condition{if} clause define \condition{sensorFaults}, and the last three conditions define \condition{trackingPilotCommands}. The \textsl{Not Active} fragment is defined in the \timing{until()} clause. Finally, the first two conditions in the \Response{} partially define the \response*{controlObjectives} fragment.

The natural-language version of \textsf{UC5\_R\_1}  lists three control objectives: \textit{``settling time, overshoot, and steady state error''}, each of which we defined in one child requirement. For example, \textsf{UC5\_R\_1.1} partially defines \condition{controlObjectives} (\textsl{Settling Time}), the other two control objectives are captured by \textsf{UC5\_R\_1.2} and \textsf{UC5\_R\_1.3}, respectively. This approach simplified each child requirement, but increased the difficulty of updating the repeated fragments.

Both \textsf{UC5\_R\_1} and \textsf{UC5\_R\_3} depend on \textsl{Sensor Faults} and \textsl{Tracking Pilot Commands} (Fig.\ref{fig:r1r3Dependancy}) meaning that the fragments' definitions are duplicated four times over the children of \textsf{UC5\_R\_1} and \textsf{UC5\_R\_3} (\textsf{UC5\_R\_3} only has one child).
The children also duplicate the definition of \textsl{Active} and \textsl{Not Active} eight times. 

In the rest of this section, we describe how we use the \textsc{Extract Requirement} refactoring on the five fragments that \textsf{UC5\_R\_1} and \textsf{UC5\_R\_1.1} depend on: \textsl{Sensor\ Faults, Control\ Objectives, Tracking\ Pilot\ Commands, Active, and Not\ Active}. We remind the reader that we defer checking that a requirement's behaviour has not changed after refactoring, until Step~\ref{sec:step4}.

\paragraph{Sensor Faults}~\\
Applying \textsc{Extract Requirement} to \textsl{Sensor Faults} was relatively easy, because its complete definition is simply repeated across several requirements. The detailed specification was extracted to a new requirement
\textsf{SENSOR\_FAULTS}: 
\begin{center}\condition{when\tnbs{}(sensorValue(S)\tnbs{}$>$\tnbs{}nominalValue+R)\tnbs{}|\tnbs{}(sensorValue(S)\tnbs{}$<$\tnbs{}nominalValue-R)\tnbs{}|\tnbs{}(sensorValue(S)=null)} \component{Controller} \texttt{shall} \response{SensorFaults}
\end{center}

\noindent After creating the new requirement, the original repeated definitions were replaced with a reference to this new requirement. For example requirements \textsf{UC5\_R\_1}  and \textsf{UC5\_R\_1.1} become: 
\begin{center}
\textsf{UC5\_R\_1} $=$ \condition{if\tnbs{}(SENSOR\_FAULTS\tnbs{}\&\tnbs{}(trackingPilotCommands))} \component{Controller} \texttt{shall} \response{(controlObjectives)}

\textsf{UC5\_R\_1.1} $=$ \condition{when\tnbs{}(diff(r(i),y(i))\tnbs{}\textgreater{}\tnbs{}E)\tnbs{}if(SENSOR\_FAULTS\tnbs{}\&\tnbs{}(pilotInput\tnbs{}=\tnbs{}\textgreater{}\tnbs{}setThrust=V2)\tnbs{}\&\tnbs{}(observedThrust=V1))} \component{Controller} \texttt{shall} \timing{until\tnbs{}(diff(r(i),y(i))$<$e)} \response{(settlingTime\tnbs{}\textgreater{}=\tnbs{}0)\tnbs{}\&\tnbs{}(settlingTime\tnbs{}\textless{}=\tnbs{}settlingTimeMax)\tnbs{}\&\tnbs{}observedThrust=V2}
\end{center}

\paragraph{Control Objectives}~\\
Applying \textsc{Extract Requirement} to \textsl{Control Objectives} was a little more difficult than for \textsl{Sensor Faults}, because its definition was spread over several children, but was still achievable. As previously mentioned, each of \textsf{UC5\_R\_1} 's child requirements (\textsf{UC5\_R\_1.1}, \textsf{UC5\_R\_1.2}, and \textsf{UC5\_R\_1.3}) contains part of the definition of \response*{controlObjectives}.

Each of the part-definitions of \response*{controlObjectives} was extracted and combined into a single requirement \textsf{CONTROL\_OBJECTIVES}. The part-definitions in the child requirements were replaced with a reference to the new requirement. For example \textsf{UC5\_R\_1}  and \textsf{UC5\_R\_1.1} became:
\begin{center} 
\textsf{UC5\_R\_1} $=$ \condition{if\tnbs{}(SENSOR\_FAULTS\tnbs{}\&\tnbs{}trackingPilotCommands)} \component{Controller} \texttt{shall} \response{CONTROL\_OBECTIVES}

\textsf{UC5\_R\_1.1} $=$ \condition{when\tnbs{}(diff(r(i),y(i))\tnbs{}\textgreater{}\tnbs{}E) if(SENSOR\_FAULTS\tnbs{}\&\tnbs{}(pilotInput=\textgreater{}\tnbs{}setThrust=V2)\tnbs{}\&\tnbs{}(observedThrust=V1))} \component{Controller} \texttt{shall} \timing{until\tnbs{}(diff(r(i),y(i))$<$e)} \response{CONTROL\_OBECTIVES\tnbs{}\&\tnbs{}observedThrust=V2}
\end{center}

\paragraph{Tracking Pilot Commands}~\\
The final natural-language fragment, \textsl{Tracking Pilot Commands}, was slightly more difficult to extract because its definition occurs in a requirement's \Condition{} field(\condition{pilotInput\tnbs{}=\textgreater{}\tnbs{}setThrust=V2)\tnbs{}\&\tnbs{}(observedThrust=V1)})  and its \Response{} field (\response{observedThrust=V2}).

We decided to extract the parts of this fragment that are in the \Condition{} to their own requirement \textsf{TRACKING\_PILOT\_COMMANDS}. However, we made a design decision to leave \response*{observedThrust=V2} in place, which simplifies the intuition of how extracted fragments work in \gls{fret}. For example \textsf{UC5\_R\_1} and \textsf{UC5\_R\_1.1} became:
\begin{center} 
\textsf{UC5\_R\_1} $=$ \condition{if(SENSOR\_FAULTS\tnbs{}\&\tnbs{}TRACKING\_PILOT\_COMMANDS)} \component{Controller} \texttt{shall} \response{CONTROL\_OBECTIVES}

\textsf{UC5\_R\_1.1} $=$ \condition{when\tnbs{}(diff(r(i),y(i))\tnbs{}\textgreater{}\tnbs{}E) if(SENSOR\_FAULTS \&\tnbs{}TRACKING\_PILOT\_COMMANDS} \component{Controller} \texttt{shall} \timing{until\tnbs{}(diff(r(i),y(i))$<$e)} \response{CONTROL\_OBECTIVES\tnbs{}\&\tnbs{}observedThrust=V2}
\end{center}

\begin{sloppypar}
In the refactored version of \textsf{UC5\_R\_1.1}, the references to \textsf{SENSOR\_FAULTS} and \textsf{TRACKING\_PILOT\_COMMANDS} guard
the update \response*{observedThrust=V2}: both references are interpreted as booleans that are intuitively \texttt{true} if their \Condition{} is \texttt{true}. 
Had we extracted \response*{observedThrust=V2} into the \Response{} of \textsf{TRACKING\_PILOT\_COMMANDS}, then \response*{observedThrust=V2} could happen even if \textsf{SENSOR\_FAULTS} is \texttt{false} because \textsf{TRACKING\_PILOT\_COMMANDS} could trigger the update independently.
\end{sloppypar}

\paragraph{Active and Not Active}~\\
As previously mentioned, \textsl{Active} and \textsl{Not Active} are contained within the \condition{when()} and \timing{until()} clauses of each child requirement. Because they define the triggering and stopping condition (respectively) for a requirement, these two fragments must be extracted separately to preserve their book-ending structure.

The \condition{when()} clause defines the triggering condition for the interval in which the requirement should be \textsl{Active} (\condition{when\tnbs{}(diff(r(i),y(i))$>$E)}) and is extracted into \textsf{Active}; the \timing{until()} clause defines the stopping condition for the interval (\timing{until\tnbs{}(diff(r(i),y(i))$<$e)}) and is extracted into \textsf{Not Active}. Each of these was then substituted into the relevant location in the child requirements; therefore, \textsf{UC5\_R\_1} remains the same, and \textsf{UC5\_R\_1.1} becomes:
\begin{center}
\condition{when\tnbs{}ACTIVE if\tnbs{}(SENSOR\_FAULTS\tnbs{}\&\tnbs{}TRACKING\_PILOT\_COMMANDS} \component{Controller} \texttt{shall} \timing{until\tnbs{}NOT\_ACTIVE} \response{CONTROL\_OBECTIVES\tnbs{}\&\tnbs{}observedThrust=V2}
\end{center}

\subsection{Step 3: Remove Redundant Requirements}
\label{sec:step3}

In this step we remove the redundant identical child requirements that were produced in Step~\ref{sec:step2}. 
As an example of the identical requirements, \textsf{UC5\_R\_1.1}, \textsf{UC5\_R\_1.2} and \textsf{UC5\_R\_1.3} \textit{all} now read:
\begin{center}
\condition{when\tnbs{}ACTIVE if\tnbs{}(SENSOR\_FAULTS\tnbs{}\&\tnbs{}TRACKING\_PILOT\_COMMANDS} \component{Controller} \texttt{shall} \timing{until\tnbs{}NOT\_ACTIVE} \response{CONTROL\_OBECTIVES\tnbs{}\&\tnbs{}observedThrust=V2}
\end{center}
\noindent Hence, we only need one child requirement to specify the detail of \textsf{UC5\_R\_1}.

In these cases, we simply remove all but one of the identical child requirements. This results in a cleaner set of requirements, which maintains traceability back to the natural-language version, and uses \gls{fret}'s parent-child link to maintain internal traceability between the abstract (parent) and detailed (child) versions of the \fretish{} requirements. Now, in the refactored set of requirements, a change to one of the fragments only needs an update in one place. 

\subsection{Step 4: Check the Requirements} 
\label{sec:step4}

Our final step is to check that the refactoring steps have not changed the behaviour of the requirements. This step is almost impossible with natural-language requirements, even those written in semi-structured languages, because of ambiguity in the requirements' meaning. However, we can formally compare the two versions of \fretish{} requirements, because \gls{fret} automatically translates them into temporal logic.

We use the Spot library\footnote{Spot library: \url{https://spot.lrde.epita.fr/}.} to compare the temporal logic generated by \gls{fret} for each refactored requirement with its original version. First, we abstracted away from the detailed terms in the \fretish{} requirements, replacing them with boolean variables. This was to ensure that the temporal logic produced by \gls{fret} was compatible with Spot. Then, we used \gls{fret} to recompile the newly abstracted \fretish{} statements, to get the corresponding temporal logic formulae. Finally, we used Spot to compare the two temporal logic formulae.

Out of the 30 requirements, 24 are equivalent, meaning that the refactoring has made no change to their behaviour. Of the remaining 6 requirements that are not equivalent, each of the factored requirements implies the original. These 6 requirements are all children: \textsf{UC5\_R\_1.1}, \textsf{UC5\_R\_2.1}, \textsf{UC5\_R\_5.1}, \textsf{UC5\_R\_6.1}, \textsf{UC5\_R\_9.1}, \textsf{UC5\_R\_10.1}. After refactoring, each of these requirements contains the full definition of \textsl{Control Objectives}; whereas, before refactoring, they only contained a part-definition, hence implication holds.

For example, the \Response{} in the original version of \textsf{UC5\_R\_1.1} defines the part of the \textsl{Control Objectives} fragment relating to settling time: \response*{(settlingTime >= 0)\tnbs{}\&\tnbs{}(settlingTime <= settlingTimeMax)}. Whereas, in the refactored version, the \Response{} defines all three parts of the fragment: 
\begin{center}
\response*{(settlingTime\tnbs{}>=\tnbs{}0)\tnbs{}\&\tnbs{}(settlingTime\tnbs{}<=\tnbs{}settlingTimeMax)\tnbs{}\&\tnbs{}(overshoot\tnbs{}>=\tnbs{}0)\tnbs{}\&\tnbs{}(overshoot\tnbs{}<=\tnbs{}overshootMax)\tnbs{}\&\tnbs{}(steadyStateError\tnbs{}>=\tnbs{}0)\tnbs{}\&\tnbs{}(steadyStateError\tnbs{}<=\tnbs{}steadyStateErrorMax)}
\end{center}
The refactored version of the requirement implies the original, because the refactored version's \Response{} covers the original's.
The behaviour of many requirements has merged into one, but overall the required behaviour is the same.

\begin{table}[t]
\centering
\begin{tabular}{c|c|c|c}
\multirow{2}{*}{\textbf{Parent ID}} & \multirow{2}{*}{\textbf{\textnumero~of Dependencies}} & \multicolumn{2}{c}{\textbf{\textnumero~of  Child Requirements}} \\
\cline{3-4}
 &  & \textbf{Before Refactoring} & \textbf{After Refactoring} \\
\hline \hline
 
\textsf{UC5\_R\_1}  & 3 & 3 & 1 \\ \hline
\textsf{UC5\_R\_2}  & 3 & 3 & 1 \\ \hline
\textsf{UC5\_R\_3}  & 3 & 1 & 1 \\ \hline
\textsf{UC5\_R\_4}  & 3 & 1 & 1  \\ \hline
\textsf{UC5\_R\_5}  & 3 & 3 & 1  \\ \hline
\textsf{UC5\_R\_6}  & 3 & 3 & 1 \\ \hline
\textsf{UC5\_R\_7}  & 3 & 1 & 1  \\ \hline
\textsf{UC5\_R\_8}  & 3 & 1 & 1  \\ \hline
\textsf{UC5\_R\_9}  & 3 & 3 & 1  \\ \hline
\textsf{UC5\_R\_10} & 3 & 3 & 1  \\ \hline
\textsf{UC5\_R\_11} & 3 & 1 & 1  \\ \hline
\textsf{UC5\_R\_12} & 3 & 1 & 1 \\ \hline
\textsf{UC5\_R\_13} & 1 &  2 & 2  \\ \hline
\textsf{UC5\_R\_14} & 1 &  2 & 2 \\  
\hline \hline
 \multicolumn{2}{c|}{\textbf{Total Child Requirements}} & 28 & 16
\end{tabular}
\caption{The number of fragment dependencies for each parent requirement, and the number of child requirements before and after refactoring. Note: this does not include \textsl{Active} and \textsl{Not Active}, because they only occur in the child requirements. \label{tab:dependancies}}
\end{table}

\section{Analysis}
\label{sec:analysis}

In this section we compare our original set of requirements to our refactored set of requirements. The original requirements set contains dependencies that always lead to duplication of information. 
To quantify the impact of our refactoring approach, we analyse the original and the refactored requirements sets, and compare: the total number of requirements, and the number of repetitions of a fragment's definition.

\begin{table}[t]
\centering
\begin{tabular}{c|l|c|c}
\multirow{2}{*}{\textbf{ID}} & \multirow{2}{*}{\textbf{Fragment Name}}  & \multicolumn{2}{c}{ \textbf{\textnumero~of  (Re)Definitions}} \\
\cline{3-4}
 &  & \textbf{Before Refactoring} & \textbf{After Refactoring} \\
\hline \hline
 
F1 & \textsl{Sensor Faults} 						& 8	 & 1 \\ \hline
F2 & \textsl{Tracking Pilot Commands} 			& 13 & 1 \\ \hline
F3 & \textsl{Control Objectives} 				& 18 & 1 \\ \hline
F4 & \textsl{Regulation Of Nominal Operation} & 14 & 1 \\ \hline
F5 & \textsl{Operating Limit Objectives} 		& 6	 & 1 \\ \hline
F6 & \textsl{Mechanical Fatigue} 				& 8	 & 1 \\ \hline
F7 & \textsl{Low Probability Hazardous Events}	& 8	 & 1 \\  
\hline \hline
F8 & \textit{Active}							& 28 & 1 \\ \hline
F9 & \textit{Not Active}						& 28 & 1 \\
\hline \hline
 \multicolumn{2}{c|}{\textbf{Total (Re)Definitions}} & 132 & 9
\end{tabular}
\caption{The number of times each fragment's definition occurs in a child requirement. Note that this table includes \textsl{Active} and \textsl{Not Active}. \label{tab:fragmentRepeats} }
\end{table}

Table~\ref{tab:dependancies} shows, for each parent requirement, upon how many fragments it depends, and how many child requirements it has both before and after refactoring. Note, the parent requirements do not depend on \textsl{Active} and \textsl{Not Active}. The number of dependencies remains constant: understanding a requirement still depends on understanding the definition of its fragments. However, the number of child requirements has been reduced, lowering the total number of requirements.

In total, we removed 12 child requirements, reducing the total from 28 down to only 16. The addition of 9 requirements, to define the fragments (\S\ref{sec:step2}), mean that the total number of requirements dropped from 42 to 39. 
The requirements set contains 14 parent requirements (rephrased to refer to the fragments), 16 
child requirements, and the 9 fragments. 

After refactoring, only one child is needed to define the detail of most of the parent requirements, because the fragments have been extracted and merged. The exception to this is \textsf{UC5\_R\_13} and \textsf{14}, which specify the system's switching behaviour between two different modes; hence they have two child requirements. 

Before refactoring, each fragment was redefined in each child requirement that used it. After  refactoring, each fragment is defined in a single requirement.  
Table~\ref{tab:fragmentRepeats} shows the number definitions of each fragment (including \textsl{Active} and \textsl{Not Active}). We reduced the number of duplicated definitions of fragments by 123: from 132 before refactoring to only 9 afterwards.

\section{Discussion}
\label{sec:discussion}
In this section we discuss how to incorporate refactoring  into \gls{fret}, and how our \fretish{} requirements compare to others in the literature.

\paragraph{User View of Extending \gls{fret}:}

In \S\ref{sec:example} we described how we \textit{manually} refactored and checked our requirements, but automatic support in \gls{fret} is preferable because it would be quicker and less error-prone. \gls{fret} should enable the user to select the parts of a requirement (possibly in different fields) that are the target for refactoring; then present the user with the \textit{applicable} refactorings, and allow them to select the (existing or new) destination requirement for the refactoring. The formal comparison of the original and refactored requirements should happen automatically, and the results should be shown to the user. 

It is also important to maintain the relatively easy readability of \fretish{}. 
In \S\ref{sec:example}, the referenced requirement names were always in block capitals, to make them obvious. This, or similar, should be supported in \gls{fret}; additionally, a requirement should have a `Depends' ID link, to document its dependencies. Also, \fretish{} notation should make it clear to the user that \condition{var=1} in the \Condition{} is a boolean comparison, but \response*{var=1} in the \Response{} is an update.
%

\paragraph{Technical View of Extending \gls{fret}:}

The `Depends' ID link, mentioned above, would also help the technical implementation of our proposals. 
As previously mentioned, \gls{fret} interprets the references to other requirements as boolean variables. But with a `Depends' ID link, \gls{fret} would know which requirements were being referenced. This link would facilitate translating requirements with dependencies to temporal logic, and to both CoCoSim~\cite{bourbouh2020cocosim} and CoPilot~\cite{dutle2020requirements}.

A simple route to updating \gls{fret}'s translations to cope with requirements that have dependencies is to simply merge the requirement and its dependencies. This approach also supports the implementation of \textsc{Inline Requirement} (\S\ref{sec:inline}). As previously mentioned, merging the \Condition{} and \Response{} fields is relatively simple; but for the \Scope{} or \Component{} fields, or if there is a conflict between the \Timing{} fields, the user may have to choose which to keep. However, this would reintroduce the repetition of definitions in the translation output. A more sophisticated approach would carry the refactoring relationship to the generated conditions, but we leave developing this as future work.

\paragraph{Requirements in the Wild:}
Our requirements set was developed in collaboration with our industrial partner~\cite{farrell2021fretting}. To the best of our knowledge, this is the first set of \fretish{} requirements that have been constructed alongside an industrial partner for a system that is still under development~\cite{farrell2022refactoring}. Related work generally present \fretish{} requirements for pre-existing example applications or conceptual systems \cite{mavridou2020ten,bourbouh2021integrating,giannakopoulou2021automated}, apart from a set of \fretish{} requirements for a robotic system \cite{farrell2022} which was constructed alongside developers of an academic prototype.

Our natural-language requirements seem to have much more repetition than in related work, though~\cite{giannakopoulou2021automated} does contain some. The natural-language version of our requirements contained repetitions; and, because we opted for a one-to-one mapping to maintain traceability, so too did our \fretish{} requirements.
This is the reality of encountering requirements for systems in-the-wild and was observed in recent work \cite{deshpande2019data}.

\section{Conclusion}
\label{sec:conclusion}

This paper adapts four code refactorings for requirements written in \fretish{}, the input language of \gls{fret}. 
We take inspiration from prior work on refactoring natural-language requirements~\cite{ramos_improving_2007}, which we extend by exploiting the temporal logical translations of \fretish{} requirements to check that refactoring steps do not alter the meaning of requirements.

In previous work, we derived \fretish{} requirements for an aerospace engine controller, from natural-language requirements and test cases supplied by our industrial partner~\cite{luckcuck2021verifiable,farrell2021fretting}.
We used our adapted \textsc{Extract Requirement} refactoring to extract duplicated definitions of concepts that were repeated in the natural-language requirements, which we call \textit{fragments}, into their own requirement. This breaks the link between a requirement's dependencies and the duplication of definitions.

Despite creating 9 new requirements, the total dropped by 3. Crucially, the number of duplicated definitions dropped by 132; each of the duplicated definitions is now only stated once. Further, the requirements themselves are now easier for a user to read, reducing the cognitive overhead in understanding them.
Formally comparing the temporal logic versions of original and refactored requirements required some manual effort, but we are confident that future work on incorporating refactoring into \gls{fret} can mechanise these checks. 
Examining our approach's scalability and applicability is ongoing work within our project.

\gls{fret} supports an integrated approach to verification, by translating \fretish{} requirements into temporal logic, and the input languages of other verification tools (CoCoSim and CoPilot). Additional future work is to examine the impact of our proposed refactoring steps on these translations.


\bibliographystyle{plain}
\bibliography{refactoring.bib}

\end{document}